\documentclass[
 reprint,
nofootinbib,
 amsmath,amssymb,
 aps,
prl,
preprintnumbers
]{revtex4-2}

\usepackage{tablefootnote}

\usepackage{graphicx}
\usepackage{dcolumn}
\usepackage{bm}
\usepackage{float}
\usepackage{comment}

\usepackage[dvipsnames]{xcolor}

\newcommand{\abra}{ABRA-10cm}

\begin{document}

\title{High-Frequency Gravitational Wave Search with ABRACADABRA-10\,cm}


\newcommand{\affMIT}{1}
\newcommand{\affCERN}{2}
\newcommand{\affFermiTheory}{3}
\newcommand{\affKavli}{4}
\newcommand{\affUNC}{5}
\newcommand{\affTUNL}{6}
\newcommand{\affToronto}{7}
\newcommand{\affBCTP}{8}
\newcommand{\affLBNLTheory}{9}
\newcommand{\affUCBPhysics}{10}
\newcommand{\affLBNLPhysics}{11}

\newcommand{\SMAuthorList}{%
Kaliro\"e~M.~W.~Pappas$^{\affMIT}$, Jessica T. Fry$^{\affMIT}$, Valerie Domcke$^{\affCERN}$, Sung Mook Lee$^{\affCERN}$, Sabrina Cheng$^{\affMIT}$, Arianna Colón Cesaní$^{\affMIT}$, Joshua W. Foster$^{\affFermiTheory,\affKavli}$, Reyco Henning$^{\affUNC,\affTUNL}$, Yonatan Kahn$^{\affToronto}$, Jonathan L. Ouellet$^{\affMIT}$, Nicholas L. Rodd$^{\affBCTP,\affLBNLTheory}$, Benjamin R. Safdi$^{\affBCTP,\affLBNLTheory}$, Chiara P. Salemi$^{\affUCBPhysics,\affLBNLPhysics}$, Inoela Vital$^{\affMIT}$, and Lindley Winslow$^{\affMIT}$%
}

\newcommand{\SMAffiliationList}{%
\small
$^{\affMIT}$Laboratory for Nuclear Science, Massachusetts Institute of Technology, Cambridge, MA 02139\\
$^{\affCERN}$Theoretical Physics Department, CERN, 1 Esplanade des Particules, CH-1211 Geneva 23, Switzerland\\
$^{\affFermiTheory}$Astrophysics Theory Department, Theory Division, Fermilab, Batavia, IL 60510, USA\\
$^{\affKavli}$Kavli Institute for Cosmological Physics, University of Chicago, Chicago, IL 60637, USA\\
$^{\affUNC}$Department of Physics and Astronomy, University of North Carolina, Chapel Hill, Chapel Hill, NC, 27599\\
$^{\affTUNL}$Triangle Universities Nuclear Laboratory, Durham, NC 27710\\
$^{\affToronto}$Department of Physics, University of Toronto, Toronto, ON M5S 1A7, Canada\\
$^{\affBCTP}$Berkeley Center for Theoretical Physics, University of California, Berkeley, CA 94720\\
$^{\affLBNLTheory}$Theoretical Physics Group, Lawrence Berkeley National Laboratory, Berkeley, CA 94720\\
$^{\affUCBPhysics}$Department of Physics, University of California, Berkeley, CA 94720\\
$^{\affLBNLPhysics}$Physics Division, Lawrence Berkeley National Laboratory, Berkeley, CA 94720%
}

\author{Kaliro\"e~M.~W.~Pappas}
\email{kaliroe@mit.edu}
\affiliation{Laboratory for Nuclear Science, Massachusetts Institute of Technology, Cambridge, MA 02139}

\author{Jessica~T.~Fry}
\affiliation{Laboratory for Nuclear Science, Massachusetts Institute of Technology, Cambridge, MA 02139}

\author{Valerie Domcke}
\affiliation{Theoretical Physics Department, CERN, 1 Esplanade des Particules, CH-1211 Geneva 23, Switzerland}

\author{Sung Mook Lee}
\affiliation{Theoretical Physics Department, CERN, 1 Esplanade des Particules, CH-1211 Geneva 23, Switzerland}

\author{Sabrina~Cheng}
\affiliation{Laboratory for Nuclear Science, Massachusetts Institute of Technology, Cambridge, MA 02139}

\author{Arianna~Colón~Cesaní}
\affiliation{Laboratory for Nuclear Science, Massachusetts Institute of Technology, Cambridge, MA 02139}

\author{Joshua~W.~Foster}
\affiliation{Astrophysics Theory Department, Theory Division, Fermilab, Batavia, IL 60510, USA}
\affiliation{Kavli Institute for Cosmological Physics, University of Chicago, Chicago, IL 60637, USA}

\author{Reyco~Henning}
\affiliation{Department of Physics and Astronomy, University of North Carolina, Chapel Hill, Chapel Hill, NC, 27599}
\affiliation{Triangle Universities Nuclear Laboratory, Durham, NC 27710}

\author{Yonatan~Kahn}
\affiliation{Department of Physics, University of Toronto, Toronto, ON M5S 1A7, Canada}

\author{Jonathan~L.~Ouellet}
\affiliation{Laboratory for Nuclear Science, Massachusetts Institute of Technology, Cambridge, MA 02139}

\author{Nicholas~L.~Rodd}
\affiliation{Berkeley Center for Theoretical Physics, University of California, Berkeley, CA 94720}
\affiliation{Theoretical Physics Group, Lawrence Berkeley National Laboratory, Berkeley, CA 94720}

\author{Benjamin~R.~Safdi}
\affiliation{Berkeley Center for Theoretical Physics, University of California, Berkeley, CA 94720}
\affiliation{Theoretical Physics Group, Lawrence Berkeley National Laboratory, Berkeley, CA 94720}

\author{Chiara~P.~Salemi}
\affiliation{Department of Physics, University of California, Berkeley, CA 94720}
\affiliation{Physics Division, Lawrence Berkeley National Laboratory, Berkeley, CA 94720}

\author{Inoela~Vital}
\affiliation{Laboratory for Nuclear Science, Massachusetts Institute of Technology, Cambridge, MA 02139}

\author{Lindley~Winslow}
\email{lwinslow@mpp.mpg.de}
\affiliation{Laboratory for Nuclear Science, Massachusetts Institute of Technology, Cambridge, MA 02139}

\begin{abstract}
 High-frequency gravitational waves (HFGWs), above 10 kHz, promise a
clean probe of new physics, largely free of the astrophysical
backgrounds that complicate lower-frequency searches. Axion
detectors, which search for axion dark matter via its coupling to
electrodynamics in a strong magnetic field, should also be sensitive
to HFGWs. We present the first dedicated search for HFGWs, using a
modified ABRA-10cm axion detector, ABRA-GW, that runs simultaneously
with a conventional axion search. ABRA-GW opens the 10\,kHz--5\,MHz
band to HFGW searches and performs the first transient search by an
axion experiment, targeting primordial black hole (PBH) mergers.
Axion sensitivity is unaffected by the added gravitational-wave
channel, and HFGW sensitivity matches theoretical expectations. This
work demonstrates the broad physics reach of axion detectors and
represents a first step toward a potential HFGW discovery.   
\end{abstract}

\maketitle
\preprint{FERMILAB-PUB-25-0306-T}

\section{\label{sec:level1}Introduction}

High-frequency gravitational waves (HFGWs), above 10\,kHz, offer a possible new window into the universe at both its smallest and largest scales. Signals in this band fall into three broad classes~\cite{Aggarwal:2025noe}: stochastic signals originating in the early universe, transient signals such as the inspirals of primordial black holes (PBHs), and persistent, coherent signals such as those from the annihilation of axions in superradiant clouds around black holes. LIGO and other interferometry-based experiments have detected gravitational waves (GWs) with sensitivity up to around 10\,kHz, which marks the highest frequency expected for known compact-object mergers, since the frequency of a merger signal scales fundamentally with the size of the emitting object~\cite{Aggarwal:2025noe, Bartolo_2016}. GWs above 10\,kHz would therefore be a smoking-gun signal of new physics. Interferometry-based GW detectors are generally not optimized for high frequencies, and as interest in this band grows, new detection methods are increasingly being explored with axion experiments; for a review, see Ref.~\cite{Aggarwal:2025noe}.

The ABRACADABRA-10\,cm (\abra) experiment was designed as a first-of-its-kind low-mass axion detector, achieving sensitivity in the mass range of 0.3 to 8\,neV, with world-leading sensitivity down to $g_{a \gamma \gamma} < 3.2 \times 10 ^{-11}\,\mathrm{GeV}^{-1}$ \cite{Kahn_2016,2019PRL,Ouellet_2019,2021PRL}. Axions are a leading DM candidate~\cite{PhysRevD.94.075001,PhysRevD.73.023505}. They are expected to couple weakly to electromagnetism, modifying Maxwell's equations and sourcing small effective currents in strong magnetic fields, which opens a path to detection. \abra~demonstrated a new detection technique using a toroidal magnet read out inductively, enabling searches for low-mass axions ($m \ll 1\,\mu{\rm {eV}}$)~\cite{PhysRevLett.112.131301, Kahn_2016}, as opposed to cavity-based experiments where the detectable axion wavelength is tied to the cavity volume~\cite{Sikivie_2021,Sikivie_detection_ratea,Cavity1,Cavity2}; see Refs.~\cite{ADMX:2024xbv,HAYSTAC:2024jch,Kim:2022hmg, Adams:2022pbo} for recent results. The success of \abra, followed by SHAFT~\cite{Gramolin_2020}, opened the path to a significant increase in axion sensitivity for similar instruments with upgraded hardware. To realize this, the \abra~and DMRadio Pathfinder experiments joined forces to launch the DMRadio program to search for GUT-scale axions~\cite{Brouwer_2022_GUT, dmradio_cryo,Brouwer_2022,dmradiocollaboration2023electromagneticmodelingsciencereach}.

In this work, we demonstrate that low-mass axion experiments can simultaneously search for axions and HFGWs, presenting the first such combined search with this technology.
In particular, we use a modified \abra~instrument, which we call ABRA-GW, to open up the 10\,kHz to 5\,MHz band to HFGW searches for the first time. GWs passing through the instrument produce an electromagnetic signal analogous to the axion signal~\cite{Gertsenshtein,GertsenshteinPt2,PhysRevD.37.1237}, allowing this search to be conducted simultaneously with the standard axion search; we verify that adding the GW channel introduces no additional noise, leaving the axion sensitivity intact. We then benchmark the GW search and find sensitivity in line with theoretical predictions, providing experimental justification for the enormous sensitivity that future instruments such as DMRadio-GUT will have to HFGWs in this window. Finally, using a basic PBH merger waveform, we perform a search for transient signals---the first such analysis performed with an axion experiment.

\section{Detection method}
Both axions and GWs modify Maxwell's equations to produce an effective current, referred to as axion~\cite{Original_Wilczek} and GW~\cite{PhysRevLett.129.041101,Berlin:2021txa} electrodynamics, respectively. These same equations are also responsible for the Primakoff effect for axions and the analogous inverse Gertsenshtein effect \cite{Gertsenshtein,GertsenshteinPt2,PhysRevD.37.1237} for GWs. Denoting the flat spacetime metric by $\eta_{\mu \nu}$, a GW $h_{\mu \nu}$ modifies this to $g_{\mu \nu} = \eta_{\mu \nu} + h_{\mu \nu}$ and is described by two polarizations $h^{+}$ and $h^{\times}$. The perturbation to the metric results in an effective current,
\begin{equation}
j_\mathrm{eff}^{\mu} \equiv \partial_{\nu} \left( -\frac{1}{2} h^\alpha_\alpha F^{\mu \nu} + F^{\mu \alpha}h_{\alpha}^{\nu} - F^{\nu \alpha}h^{\mu}_{\alpha}\right)\!.
\end{equation}
In the presence of a large magnetic field, the GW induces a current, much like the axion.

The derivation of the GW signal in \abra~was performed in Refs.~\cite{PhysRevLett.129.041101,domcke2024symmetries}. At the heart of \abra~is a toroidal magnetic field. We take the toroid to have inner radius $R$, width $a$, height $H$, and peak field value $B_{\mathrm{max}}$ (see Fig.~\ref{fig:setup2}). Inside the toroid the DC magnetic field is azimuthal and varies with the radial distance, $\rho$, as $\mathbf{B}_0 =  B_{\mathrm{max}} (R/\rho) \hat{\mathbf{e}}_{\phi}$. Outside, the field vanishes. The GW interacts with $\mathbf{B}_0$ to generate an oscillatory magnetic field at the angular frequency of the GW, $\omega$. The induced field is determined by the Biot-Savart law,
\begin{equation}
\mathbf{B}_h(\mathbf{r}') = \int_{V_B} \hspace{-0.25cm}d^3 \mathbf{r}\, \frac{\mathbf{j}_{\textrm{eff}}(\mathbf{r}) \times (\mathbf{r}' - \mathbf{r})}{4\pi |\mathbf{r}' - \mathbf{r}|^3}.
\end{equation}
The integral is taken over the volume of the toroid's magnetic field, $V_B$. Like the axion searches of \abra, we search for this induced magnetic field, oscillating with the GW frequency, as our signal.

The HFGW magnetic field can be detected from the changing magnetic flux through a pickup loop placed in the center of the toroid, as shown in Fig.~\ref{fig:setup2}. Integrating over the pickup loop area, $A_{\ell}$, the flux is given by
\begin{equation}
\Phi_h = \int_{A_\ell} \hspace{-0.2cm} d\mathbf{A}'\cdot \mathbf{B}_h (\mathbf{r'}),
\end{equation}
which induces a current in the pickup loop that is read out with quantum current sensors. Unlike the axion, which generates an azimuthally symmetric field, the magnetic field induced by a GW has an approximately dipolar structure. This structure results in a weaker coupling to an azimuthally-symmetric pickup. Since \abra~operates in the magnetostatic limit, $\omega L \ll 1$,\footnote{In general, the magnet is not rigid in the presence of the GW, leading to additional induced magnetic fields of ${\cal O}(h B_0)$~\cite{Berlin:2023grv,Domcke:2024mfu}. Since the \abra~pickup loop is placed in a region where $B_0 \simeq 0$, this does not affect the results presented here.} with $L = \{R, a, H\}$ denoting the typical scales of the detector, we can expand the predicted signal in powers of $\omega L$. For a circular pickup loop, the leading order contribution, ${\cal O}(\omega^2 L^2)$, cancels and the signal is exclusively sourced by the $h^\times$ polarization~\cite{domcke2024symmetries},
\begin{equation}
\Phi_c = \frac{i e^{-i \omega t}}{48 \sqrt{2}} h^{\times} \omega^3 B_{\mathrm{max}} \pi r^2 a R (a + 2R) \sin^2\theta_h,
\end{equation}
where $\theta_h$ indicates the direction of the incoming GW relative to the symmetry axis of the detector. However, if a semicircular figure-8 loop (see GWC and GWP in Fig.~\ref{fig:setup2}) is used instead, then the dipole-structured signal is able to couple to terms up to $\omega^2$. The figure-8 pickup additionally couples to both the $h^{\times}$ and $h^{+}$ components of polarization. The flux for a figure-8 pickup is approximated as~\cite{domcke2024symmetries}
\begin{equation}\begin{aligned}
\Phi_8 = \; & \frac{e^{-i \omega t}}{3 \sqrt{2}} \omega^2 B_{\mathrm{max}} r^3 R \, \ln (1 + a/R) \sin\theta_h \\
&\times (h^{\times} \sin\phi_h - h^+ \cos \theta_h \cos\phi_h).
\label{eq:Phi8}
\end{aligned}\end{equation}
These equations for flux are derived in the limit that the magnet is a rectangular toroid with $H \gg a,R$. In the analysis presented here we go beyond these approximations, computing the flux on the pickup loop resulting from a GW by simulation as outlined below.

\section{\label{sec:level2}Detector}

\begin{figure}[!t]
    \centering
    \includegraphics[width=0.35\textwidth]{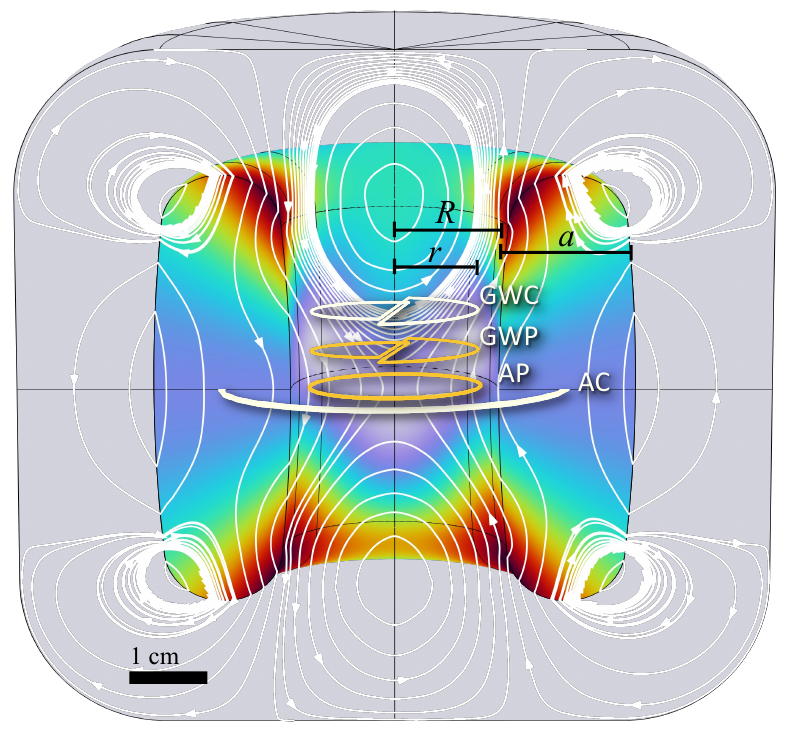}
    \caption[ABRA-GW setup]{Schematic of ABRA-GW for a wave entering from the right side of the figure. The magnitude of the effective HFGW current is shown as a color gradient (white/purple as smallest magnitude and black/red as the largest magnitude) in the volume of the toroidal magnet and the magnetic field lines resulting from the HFGW effective current are shown in white. The two orange loops in the center of the magnet are the axion pickup (AP) and GW pickup (GWP) and the white loops are the axion calibration loop (AC) and the GW calibration loop (GWC). The magnet width, $a$, is 3\,cm, the inner radius, $R$, 3\,cm, and the radius of the pickup loops, $r$, is 2.22\,cm.}
    \vspace{-0.5cm}
    \label{fig:setup2}
\end{figure}

ABRA-GW is designed to simultaneously look for axions and HFGWs. ABRA-GW was built as a modification to the original \abra~axion detector~\cite{2019PRL,Ouellet_2019,2021PRL}, which consists of a superconducting toroidal magnet with an NbTi pickup structure inside a superconducting shield, all cooled to cryogenic temperatures. Superconducting quantum interference devices (SQUIDs) were used to detect signals in the form of small-currents in the pickup structures. The magnet, pickups, and shield are cooled to $\sim$1\,K and the two-stage Magnicon DC SQUIDs are cooled to $\sim$1.3\,K using an Oxford Instruments Triton 400 dry dilution refrigerator. The pickup system for ABRA-GW is constructed with two pickups for detection: a circular pickup for axion detection, and a semicircular figure-8 pickup for HFGW detection. To first order, the axion signal does not induce a current in the figure-8 loop, and the HFGW signal does not induce a current in the circular loop as a result of low mutual inductance between the two shapes. Through simulation with COMSOL, we found a mutual inductance of $\sim$$2 \times 10 ^{-5}$\,nH between the circular and the figure-8 pickup loops (AP and GWP in Fig.~\ref{fig:setup2}), as compared to $\sim$$30$\,nH between the figure-8 pickup and calibration loops (GWP and GWC in Fig.~\ref{fig:setup2}), and $\sim$$21$~nH between the circular pickup and calibration loops (AP and AC in Fig.~\ref{fig:setup2}). This six-orders-of-magnitude difference in the mutual inductance between cross-signal measurements and same-signal measurements allows for simultaneous searches of axions and HFGWs.

The experiment was calibrated using superconducting loops of NbTi wire shaped to mimic the desired signals. For the axion signal, the original setup from the \abra~axion searches was used, consisting of a circular loop of wire with a radius of 4.5 cm running through the center of the magnet \cite{2019PRL}. An oscillating current is injected into the circular axion calibration loop and measured using the pickup structures. Similarly, an oscillating current is also injected into the HFGW calibration loop, which is in the shape figure-8 to mimic a HFGW signal. The HFGW calibration loop is located 1 cm above the figure-8 pickup loop.
A diagram of the pickups and calibration loops inside the shield containing the magnet is shown in Fig.~\ref{fig:setup2}. For each calibration signal injection, data was collected for both pickup structures to measure system response and cross correlation. Calibration data was collected before and after science data collection to ensure detector stability. The calibrations provide the detector response to a one-dimensional current in the shape of the signal, but the volumetric nature of the axion and HFGW signals was studied using simulations done with COMSOL.

The effective current resulting from a HFGW in the magnet volume was input to COMSOL using the equations found in Ref.~\cite{domcke2024symmetries}, with extensions to account for the finite height of the magnet. The full function for the effective current can be found in the supplementary material (SM); terms up to $\omega^2$ were input to COMSOL in the form of external current density. The flux on the pickup resulting from the effective current was found by integrating over the surface of the pickup area.

\section{\label{sec:level3}Data Collection and Backgrounds}

Data for the axion and HFGW searches were collected simultaneously in time as two constant streams. An AlazarTech 9870 8-bit digitizer~\cite{alazar} was used for data collection and the digitizer was locked to a Stanford Research Systems FS725 Rubidium frequency standard. The sampling rate used was 10\,MHz, giving a Nyquist frequency of 5\,MHz. We used a physical 10\,kHz high-pass filter and a 5\,MHz low-pass filter to attenuate the noise at low frequencies and prevent aliasing. We collected data for six days from December 4 to 9, 2023 on both the axion and HFGW channels, with a brief break after the first few hours as a result of memory buffer issues in the data-taking computer.

The raw data contained a few distinct background signals including a 5 MHz signal and a 13\,kHz signal, confirmed to exist in the data taken with both the magnet charged and uncharged. After using a Butterworth filter from 10 kHz to 3 MHz, the 13\,kHz signal remained as the most prominent background in the data. The digital filtering was performed after all the data had been collected.

\section{\label{sec:level4}Results}

\vspace{0.2cm} \noindent {\bf Axion Sensitivity.} The axion data for the ABRA-GW search was used to determine the potential sensitivity of the setup to axions. An axion search was not performed for this paper; the data collected could be analyzed identically to Refs.~\cite{Foster:2021ngm,2019PRL,Ouellet_2019,2021PRL} to create axion limits. 

\begin{figure}[!t]
    \centering
    \includegraphics[width=0.45\textwidth]{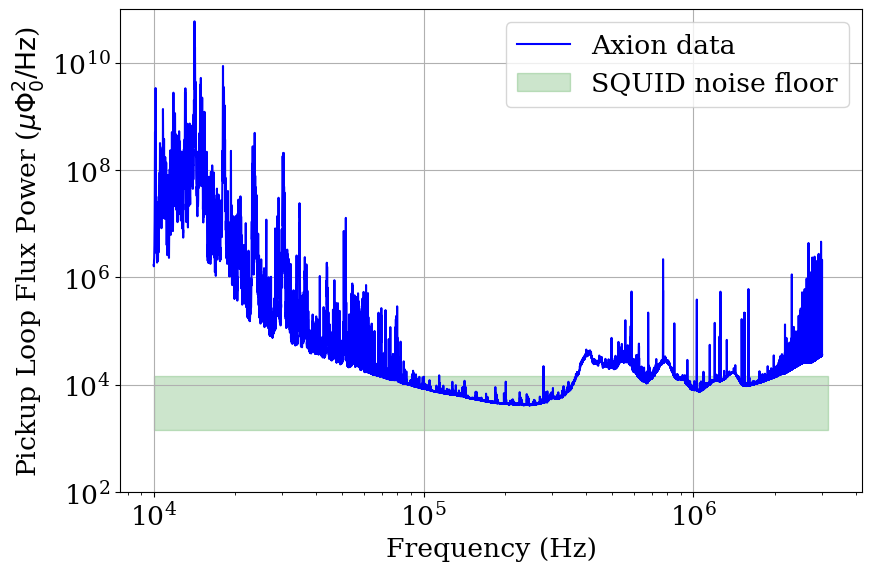}
    \vspace{-0.2cm}
    \caption[Run 6 axion sensitivity]{Axion data collection (blue) and SQUID noise floor (green). Data is shown in power on the pickup loop, with all hardware filters divided out. The SQUID noise is for a nominal 2-stage Magnicon SQUID reflecting a temperature-based range in performance from 300 mK to 4.2 K. For ABRA-GW the SQUID sat at $\sim 1.3$\,K. We observed that the axion data taken simultaneously with the HFGW collection had a noise floor that matched the SQUID noise floor at high frequency.} 
    \label{fig:axion}
    \vspace{-0.5cm}
\end{figure}

The noise on the axion search was determined by comparing the axion data power spectral density (PSD) to the SQUID noise floor PSD, similar to Ref.~\cite{2019PRL}. The SQUID noise floor is defined as the output of the SQUID when there is no connected pickup loop and therefore no direct input to the SQUID. We demonstrate with Fig.~\ref{fig:axion} that the axion sensitivity in the region of interest (100\,kHz to 2\,MHz, based on Ref.~\cite{2021PRL}) was not degraded by the HFGW search. The noise features of the spectrum are similar to ones encountered in previous \abra~campaigns, for more information on how these features are handled in an axion search see the SM of Ref.~\cite{2021PRL}. The axion data taken during the six-day-long collection period is segmented, transformed to PSDs, averaged down, and compared to the SQUID noise floor. All the data shown in the figure are corrected for the shapes of the physical high and low pass filters. The calibration data was used to transform the spectra into the equivalent input at the axion pickup loop. For the SQUID noise floor, we used data taken a few months prior with the same SQUID that was used for the axion data. Similarly to the ABRA-GW axion result, figure 5 of Ref.~\cite{Ouellet_2019} shows that first axion search of \abra~also had a noise floor close to that of the SQUID noise floor. The noise floor in Ref.~\cite{Ouellet_2019} was determined analytically instead of experimentally, as the first axion search did not take data with the SQUID disconnected.

\begin{figure}[!t]
    \includegraphics[width=0.45\textwidth]{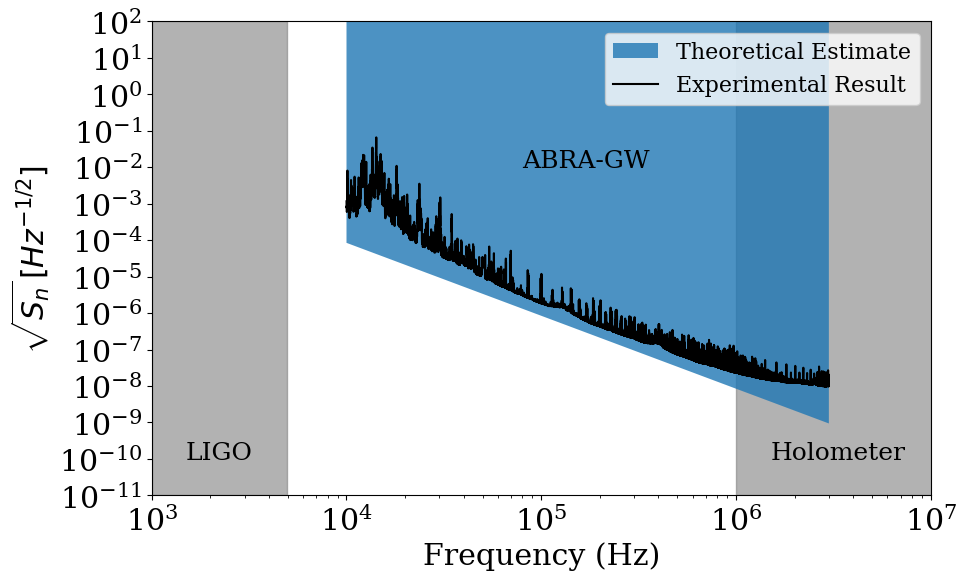}
    \vspace{-0.2cm}
    \caption[Noise Equivalent Noise Power]{Strain-equivalent noise power as theoretically estimated and as achieved by ABRA-GW. Results are purely the $h^{\times}$ polarization of the strain in the angular direction which maximizes amplitude. The data was averaged over a six-day-long period. The gray areas represent the limits achieved by the LIGO\cite{LIGO:2024kkz} and Holometer experiments\cite{Holometer:2016qoh}.} 
    \label{fig:NES}
    \vspace{-0.5cm}
\end{figure}

\vspace{0.2cm} \noindent {\bf GW Strain Sensitivity.} 
The strain-equivalent noise (SEN) is the standard metric used to compare HFGW detector technologies~\cite{Aggarwal:2025noe}, folding a detector's intrinsic noise with its response function to give an equivalent limit on GW strain, independent of the source. We assume a stationary, constant signal over the data collection period. To obtain the SEN, we used calibration results to convert digitizer voltage to pickup flux, and simulation to convert flux on the pickup to strain, taking the equations for effective current from Ref.~\cite{domcke2024symmetries} as input. We choose the GW parameters that produce the strongest signal: an incoming wave at incident angle $\theta_h = \phi_h = \pi/2$, for which the $h^\times$ term dominates by an order of magnitude, and we approximate the signal as purely $h^\times$. The theoretical calculation shown in Fig.~\ref{fig:NES} uses the characteristic DC SQUID noise floor with the same signal parameters as the simulation, with the flux on the pickup estimated using Eq.~\eqref{eq:Phi8}; further details are given in the SM.

As shown in Fig.~\ref{fig:NES}, the experimental result closely matches the theoretical expectation, confirming that ABRA-GW achieves its predicted sensitivity in operation. This is the first demonstration that an axion detector can search for HFGWs without degrading its axion sensitivity. The result is also the first limit at frequencies above those probed by LIGO O4a~\cite{LIGO:2024kkz}, sitting just above the frequency range where astrophysical sources are expected to end. The Holometer experiment sets complementary bounds above 1 MHz~\cite{Holometer:2016qoh}.

\vspace{0.2cm} \noindent {\bf Transient Search.} We can use this demonstrator to perform the first time-series transient search for HFGW. For the purposes of this analysis, we only computed one signal template of two equal-mass PBHs ($m_1 = m_2 = 0.01\,M_{\odot}$) with no eccentricity and a fixed sky position to use in our transient analysis. The analysis framework can be generalized to multiple templates, time and computing resources permitting. This template models the scenario where the PBH merger event is at some distance from the center of the detector. All parameters were chosen to maximize the strain in the merger within our frequency window (10\,kHz to 3\,MHz) and are listed in the SM. The ripple code base \cite{edwards2023ripple} was used to generate the waveforms of the two polarizations of strain for the merger. The amplitude of the signal scales inversely to the distance from the detector to the source. Our template models a merger event occurring 100 km from the center of mass of the detector, corresponding to strain $\mathcal{O}(10^{-4})$, as shown in the SM.

As a result of the 13\,kHz background signal and other unidentified noise sources, the local mean of the GW data is non-zero and a traditional matched-filter would require a more complicated implementation for our analysis. To account for this non-zero mean we use Gaussian process modeling (GPM). Once the raw data is filtered to the same frequency range as the template (10\,kHz to 3\,MHz), GPM is used to construct the marginalized likelihood for the residuals $\mathbf{r}(d)$, which are the difference between the detector data and the expected signal for a signal at distance $d$ from the detector. The test statistic is given by
\begin{equation}
\mathrm{TS}(d) = -2 \ln \frac{p(\mathbf{r}|d)}{p(\mathbf{r}|\hat{d})}
\end{equation}
for the log-marginalized likelihood of the residuals at distance $d$, where $\hat{d}$ is the maximum likelihood estimate of the signal distance. For the transient search, we allow the distance to vary in each segment of data analyzed and find the smallest distance we can exclude at a 90\% confidence level (CL) for that segment.

With a search window of $30\,\mu\mathrm{s}$, corresponding to the length of the
signal template, we use a time step of $0.4\,\mu\mathrm{s}$ to ensure the
signal will not be missed when scanning over data segments. As a demonstration,
we ran the GPM analysis on a portion of the data. For a total data collection
time $T = 518459\,\mathrm{s} \simeq 144\,\mathrm{hrs}$, we have
$N_\mathrm{process} = 1.29 \times 10^{12}$ total segments to process.
Analyzing the full dataset is beyond the scope of this work; instead, the
pathfinder analysis conducted here used a 10 ms portion of the data. We
tested the GPM analysis on the lowest noise period of the data (determined
by power in 10 ms bins, see the SM), along with an average noise period of
the data. Comparing these periods, we found that the distance which all
segments within each period excluded to at least a 90\% CL was similar
across more and less noisy time periods, as expected given the overall low
variance in the noise over the six-day collection period (see the SM). The
proof-of-principle analysis conducted here should be viewed as obtaining
conservative limits; refinements could strengthen the sensitivity.

Out of 24975 windowed data segments searched in our pathfinder analysis,
corresponding to 10 ms of collection time, all segments excluded a distance
of 46.03 km (strain $\mathcal{O}(10^{-4})$) to at least a 90\% CL. Given
that we expect the dominant signal contribution around the merger
frequency, which for $0.01\,M_\odot$ PBHs is $f_\text{isco} \simeq
0.2$\,MHz, this is consistent with expectation from the strain-equivalent
noise in Fig.~\ref{fig:NES}: $h_\text{sens} \sim \sqrt{S_n f_\text{isco}}
\simeq 4 \times 10^{-5}$. More details are given on determining this
distance in the SM.

\begin{figure}[!t]
    \centering
    \includegraphics[width=0.45\textwidth]{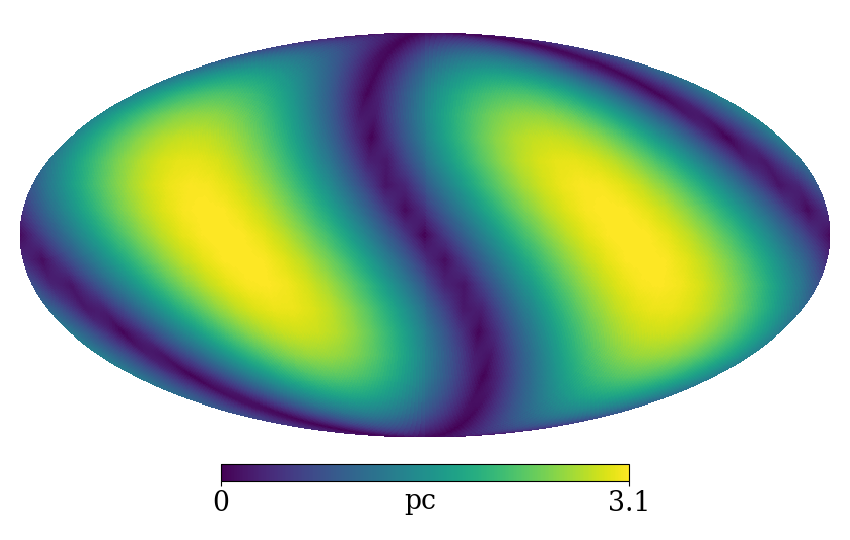}
    \vspace{-0.2cm}
    \caption[Distance Results]{Distance from detector to source for the minimum excluded amplitude found from the GPM analysis, projected for DMRadio-GUT and plotted over the sky in instrumental coordinates with the $z$-axis aligned with the center of the sky map.}
    \label{fig:DistanceBlock}
    \vspace{-0.5cm}
\end{figure}

Beyond improvements to the analysis itself, the next generation of
GUT-scale axion experiments, the DMRadio-GUT
experiment~\cite{Brouwer_2022_GUT}, will have a sensitivity much larger
than ABRA-GW. Using the results of Ref.~\cite{PhysRevLett.129.041101}, we
expect an improvement of $\simeq 2\times10^{12}$ in strain sensitivity
from ABRA-GW to DMRadio-GUT, even with this conservative analysis; if the
mechanical readout scheme of Ref.~\cite{Domcke:2024mfu} were adopted, the
sensitivity could be improved parametrically further still. Scaling our
ABRA-GW result to the sensitivity of DMRadio-GUT, we obtain a strain
sensitivity of $\mathcal{O}(10^{-16})$, corresponding to distances of
$\mathcal{O}(1)$ parsecs, comparable to the scale of the solar
neighborhood. Figure~\ref{fig:DistanceBlock} shows the DMRadio-GUT
projection for limits on the distance of the source as a function of the
source position across the sky, in coordinates where the z-axis lines up
with the z-axis of the instrument. Given the length of the data analyzed,
there are portions of the sky to which we appear to be insensitive;
however, more time analyzed would allow for movement of the detector as a
result of the rotation of the Earth, and we would achieve full sky
coverage.

\section{Conclusion}
We have demonstrated the first conversion of a broadband lumped-element
axion detector to a HFGW detector and the first time-series transient
search with an axion experiment. ABRA-GW achieved the predicted
sensitivity to HFGW without a decrease in sensitivity to axions. This
required a new detector layout capable of reading out two simultaneous
data streams, along with simulations to characterize the expected signal
and a dedicated analysis pipeline using GPM to handle the resulting
non-zero-mean backgrounds. The sensitivity to HFGWs we achieved matches
theoretical expectation, giving us the infrastructure to scale with more
sensitive experiments like DMRadio-GUT, which will reach down to strains
$\mathcal{O}(10^{-16})$, corresponding to distances of $\mathcal{O}(1)$ parsecs,
comparable to the scale of the solar neighborhood~\cite{Brouwer_2022_GUT}.

Realizing this potential will require improvements on two fronts. First,
both online and offline algorithms will need further development to
process the much larger data volumes and search spaces that future, more
sensitive searches will demand. Second, the detectors themselves must
improve. Future axion experiments can increase their HFGW sensitivity
along several complementary directions. Changing pickup geometry, as done
here with ABRA-GW, is one route. Operating multiple detectors in
coincidence is another, where correlated signals and uncorrelated noise
between detectors improve sensitivity to both PBH mergers and stochastic
or persistent signals of new physics~\cite{Foster:2020fln}. A third,
complementary strategy reinterprets the experiment's magnet itself as a
mechanical Weber bar, exploiting its response to the direct mechanical
strain of a passing GW rather than only the induced electromagnetic
current~\cite{Domcke:2024mfu}; realizing this in practice will require a different geometry than ABRA-GW's current toroidal design. Together, these approaches, improved
geometry, multi-detector coincidence, and Weber bar readout, chart a path
for axion detectors to become competitive HFGW detectors. This work
represents a first step toward that goal.

\begin{acknowledgments}
We would like to thank our \mbox{DMRadio} colleagues for useful discussions. This  research  was  supported  by  the  National  Science  Foundation  under  grant  numbers NSF-PHY-2411650, NSF-PHY-1658693,  and NSF-PHY-1806440.
The research of N.L.R. was supported by the Office of High Energy Physics of the U.S. Department of Energy under contract DE-AC02-05CH11231.
R.H. is supported by the US Department of Energy, Office of Science, Office of Nuclear Physics grants DEFG02-97ER41041 and DEFG02- 97ER41033.
B.R.S. is supported in part by the DOE award DESC0025293, and B.R.S. acknowledges support from the Alfred P. Sloan Foundation.
J. T. Fry is supported by the National Science Foundation Graduate Research Fellowship under Grant No. 2141064.
C. P. S. is supported in part by the National Science Foundation Graduate Research Fellowship under Grant No. 1122374.
This manuscript has been authored in part by Fermi Forward Discovery Group, LLC under Contract No. 89243024CSC000002 with the U.S. Department of Energy, Office of Science, Office of High Energy Physics.
\end{acknowledgments}

\bibliographystyle{utphys}
\bibliography{ref}

\appendix

\clearpage

\onecolumngrid

\begin{center}
  \textbf{\large Supplementary Material for High-Frequency Gravitational Wave Search with ABRACADABRA-10cm}\\[.2cm]
  \vspace{0.05in}
  \SMAuthorList\\[.15cm]
  \SMAffiliationList
\end{center}

\twocolumngrid
\setcounter{equation}{0}
\setcounter{figure}{0}
\setcounter{table}{0}
\setcounter{section}{0}
\setcounter{page}{1}
\makeatletter
\renewcommand{\theequation}{S\arabic{equation}}
\renewcommand{\thefigure}{S\arabic{figure}}
\renewcommand{\thetable}{S\arabic{table}}

\subsection{Strain-Equivalent Noise}

Finding the SEN experimentally is effectively the inverse process to determining the instrument response to the merger template. For the SEN we start with the digitizer noise power from the full data set,\footnote[1]{We take the case where there is no HFGW signal for the full data set without losing our ability to claim discovery by using the fact that the signals we look for are transient and lasting ${\cal O}(1\,\mu{\rm s})$. To find the SEN we rfft the data and average over the span of data collection, any transient signal we might be looking for would be averaged to zero and we lack the sensitivity to search for a stochastic signal with only one detector.} then work backwards to convert to strain. Since the GW data for ABRA-GW was taken in time ($n_\mathrm{out}(t)$), we first had to find the PSD of the time series data. The data was divided into 1\,s long segments, those segments were Fourier transformed ($\Tilde{n}_\mathrm{out}(f)$), squared, and divided by frequency step to get the PSD ($\mathcal{F}_{V}(f)$), then all the individual PSDs were averaged together ($\bar{\mathcal{F}}_{V}(f)$). The final $\bar{\mathcal{F}}_{V}(f)$ was averaged over the total collection time of six days, representing the averaged PSD of the detector output for the full run.

The input to the detector was found by multiplying $\bar{\mathcal{F}}_{V}(f)$ by the inverse of the transfer function squared, which converts flux on the pickup to voltage on the digitizer, 
\begin{equation}
    \bar{\mathcal{F}}_{V}(f) [\mathcal{T}^{-1}(f)_{\Phi \rightarrow V}]^2 = \mathrm{\mathcal{F}_{\Phi}}(f).
    \label{eq:SEN}
\end{equation}
To get the SEN, we used the COMSOL simulation described in the Detector section of the main text to convert flux on the pickup loop to strain. Using the equations for effective current from Ref.~\cite{domcke2024symmetries} as input and separating the $h^+$ and $h^\times$ terms into distinct simulations, we use Eq.~\eqref{eq:SEN} to find either $h^+$ or $h^\times$. Since the simulation results for $h^+$ and $h^\times$ are not equal, the polarizations must be handled separately. When choosing an incident angle to produce the strongest signal, the $h^\times$ term dominates by an order of magnitude for $\theta_h = \phi_h = \pi/2$. With this relation in mind, we choose to approximate the SEN for only the $h^\times$ term. With the simulation-based transfer function, we found $\bar{\mathcal{F}}_{\mathrm{strain}_{h^{\times}}}(f) = S_n(f)/2$. The factor of 2 relating $S_n(f)$ and $\bar{\mathcal{F}}_{\mathrm{strain}_{h^{\times}}}(f)$ results from us taking a two-sided PSD on the raw data. The resulting values of $\sqrt{S_n}$ as measured by ABRA-GW are shown as the black curve in Fig.~\ref{fig:NES}.

The theoretical estimate for the sensitivity is computed as follows. The GW flux is determined from Eq.~\eqref{eq:Phi8}, taking $\theta_h = \phi_h = \pi/2$. The flux at the SQUID is taken to be $\Phi_{\textrm{SQ}} = (M/L_p)\, \Phi_8$, with $M \simeq 2.52\,$nH the mutual inductance between the SQUID and pickup loop circuit, with the latter having inductance $L_p \simeq 500\,$nH. This determines how a GW signal would manifest as a flux in the SQUID; taking the determined SQUID noise we can invert this mapping to compute the SEN, obtaining $\sqrt{S_n} \simeq 8.4 \times 10^{-9}\,\textrm{Hz}^{-1/2} (f/1\,\textrm{MHz})^{-2}$, as shown in Fig.~\ref{fig:NES}. We note that this estimate is naively optimistic: as discussed in Ref.~\cite{domcke2024symmetries}, Eq.~\eqref{eq:Phi8} overestimates the true GW flux by neglecting the finite height contributions.

\subsection{Merger Signal}

All parameters used to generate the merger signal were chosen to maximize the strain of the merger within our frequency window (10\,kHz to 3\,MHz) and are listed in Table~\ref{tab:template_parameters}. The \texttt{ripple} code base \cite{edwards2023ripple} was used to generate the waveforms of the two polarizations of strain ($h^+(\omega)$ and $h^{\times}(\omega)$). \texttt{ripple} outputs the waveforms as separate polarization arrays in frequency space, which we generate from 1 Hz to 5 MHz. The arrays could in principle be generated from 10\,kHz to 5\,MHz if we intended to stay in the frequency domain, however the transition to the time domain requires either the full frequency spectrum or a very carefully taken FFT with zero-padded arrays. We transform the arrays output from ripple to the detector (mV on the digitizer) in the frequency domain before adding them together, taking an inverse FFT, and dividing by $\Delta t$.

\renewcommand{\arraystretch}{1.2}
\begin{table}[!t]
\begin{center}
\begin{tabular}{|c | c|} 
 \hline
  Merger Parameter & Value  \\
 \hline\hline
  $M_1$ &  0.01\,$M_{\odot}$\\
 \hline
  $M_2$ &  0.01\,$M_{\odot}$\\ 
 \hline
  Dimensionless spin &  0\\ 
 \hline
  Time of coalescence &  1\,ms\\ 
 \hline  
  Distance to source &  100\,km\\ 
 \hline
  Inclination &  0\\
 \hline
\end{tabular}
\end{center}
\vspace{-0.3cm}
 \caption{Ripple inputs for the merger template.}
 \label{tab:template_parameters}
 \vspace{-0.5cm}
\end{table}

The transfer function used to convert the strain output from ripple to voltage on the digitizer is in two parts: the first relies on finding the induced flux on the pickup using the analytic form of the effective current from Ref.~\cite{domcke2024symmetries} and COMSOL simulation, and the second part converts from flux on the pickup to voltage on the ADC and is found from the calibrations. 

The full function for the effective current can be found later in the SM, the external current density was input into COMSOL with terms up to $\omega^2$. The flux through the pickup as a result of the effective current was found by integrating over the surface of the pickup. The effective current for the $h^+$ and $h^{\times}$ polarization terms were simulated separately. By testing different values of the strain polarizations, we ensured that the scaling with strain was linear and additive in terms of the two polarizations. To simplify the COMSOL inputs, terms which were constant in the effective current were accounted for post-simulation. The flux on the pickup is found to be
\begin{equation}
    \Phi_{\mathrm{pickup}} = \frac{1}{\mu_0} \times h^{+/\times}(\omega) \times \omega^2 \times \mathrm{COMSOL},
\end{equation}
where here $\mu_0$ is the permeability of free space, the $h^{+/\times}(\omega)$ terms represent the waveforms for the $h^{+}$ and $h^{\times}$ components of the incoming wave and ``COMSOL" is the output of the simulation and accounts for the geometry of the experiment.

\begin{figure}[!t]
    \centering
    \includegraphics[width=0.48\textwidth]{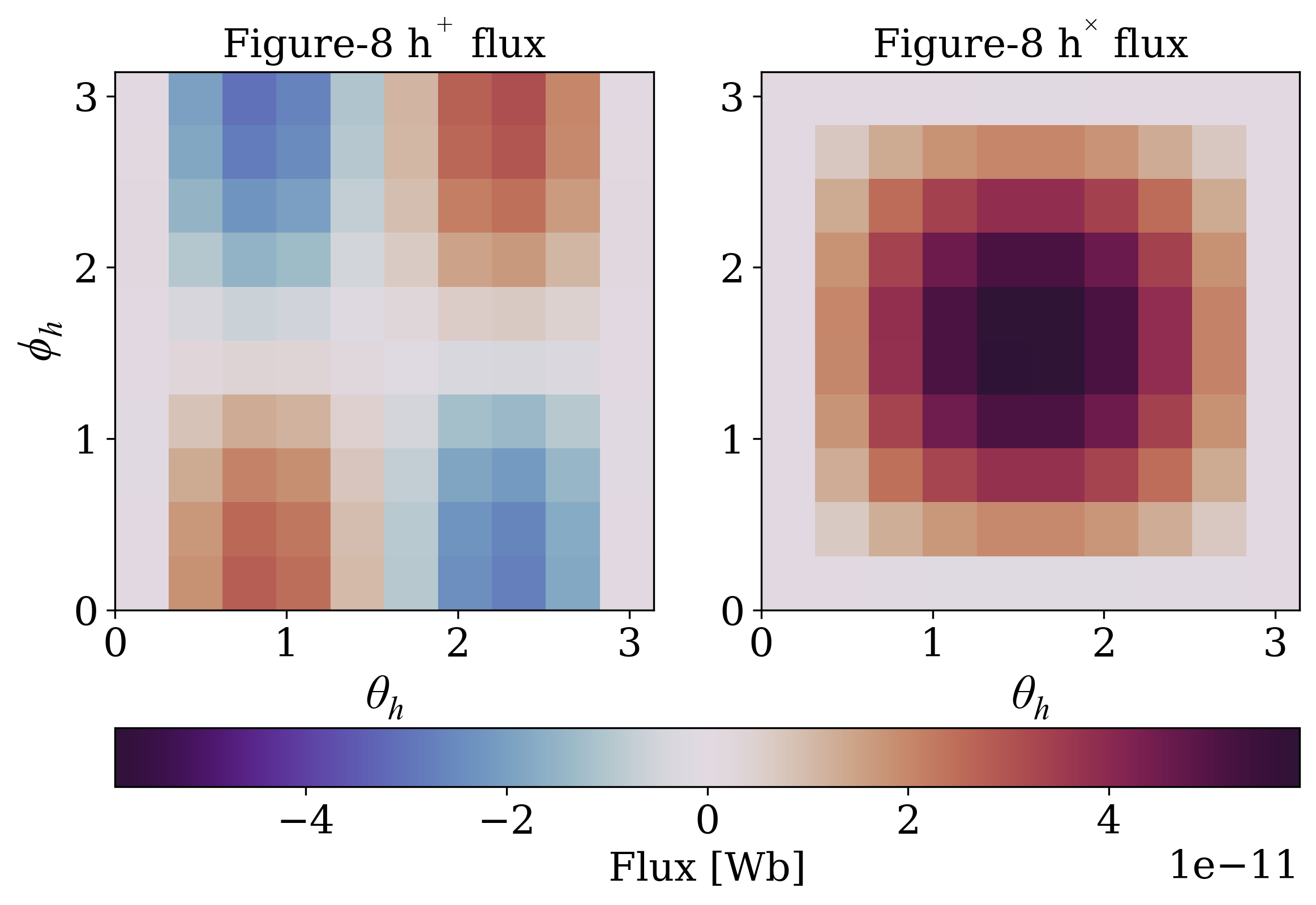}
    \vspace{-0.3cm}
    \caption[Strain polarizations in ABRA-GW]{Each square in the plots is the magnitude of the pickup-loop flux for a given pair of angles. Results are shown for flux calculated over the figure-8 area for $h^{\times} = 0$, $h^+ = 10^{-4}$ (left) and $h^{\times}$ = $10^{-4}$, $h^+ = 0$ (right).}
    \label{fig:h+hx_heatplot}
    \vspace{-0.2cm}
\end{figure}

As a result of our goal to create and run our analysis with one template, we chose only one angle for the incoming wave, which was $\pi / 2$ for both $\phi_h$ and $\theta_h$. This angle was chosen to maximize the $h^{\times}$ component, since the COMSOL results for $h^{\times}$ were on average an order of magnitude larger. Figure~\ref{fig:h+hx_heatplot} shows the COMSOL flux results for each polarization for incoming angles from 0 to $\pi$ in $\theta_h$ and from 0 to $2\pi$ in $\phi_h$. In an analogy to the axion search, the conversion of strain to flux on the pickup for the HFGW search includes similar information to the axion geometric coupling term $\mathcal{G}_V$, which connects the axion current to the current on the pickup.

\begin{figure}[!t]
    \centering
    \includegraphics[width=0.45\textwidth]{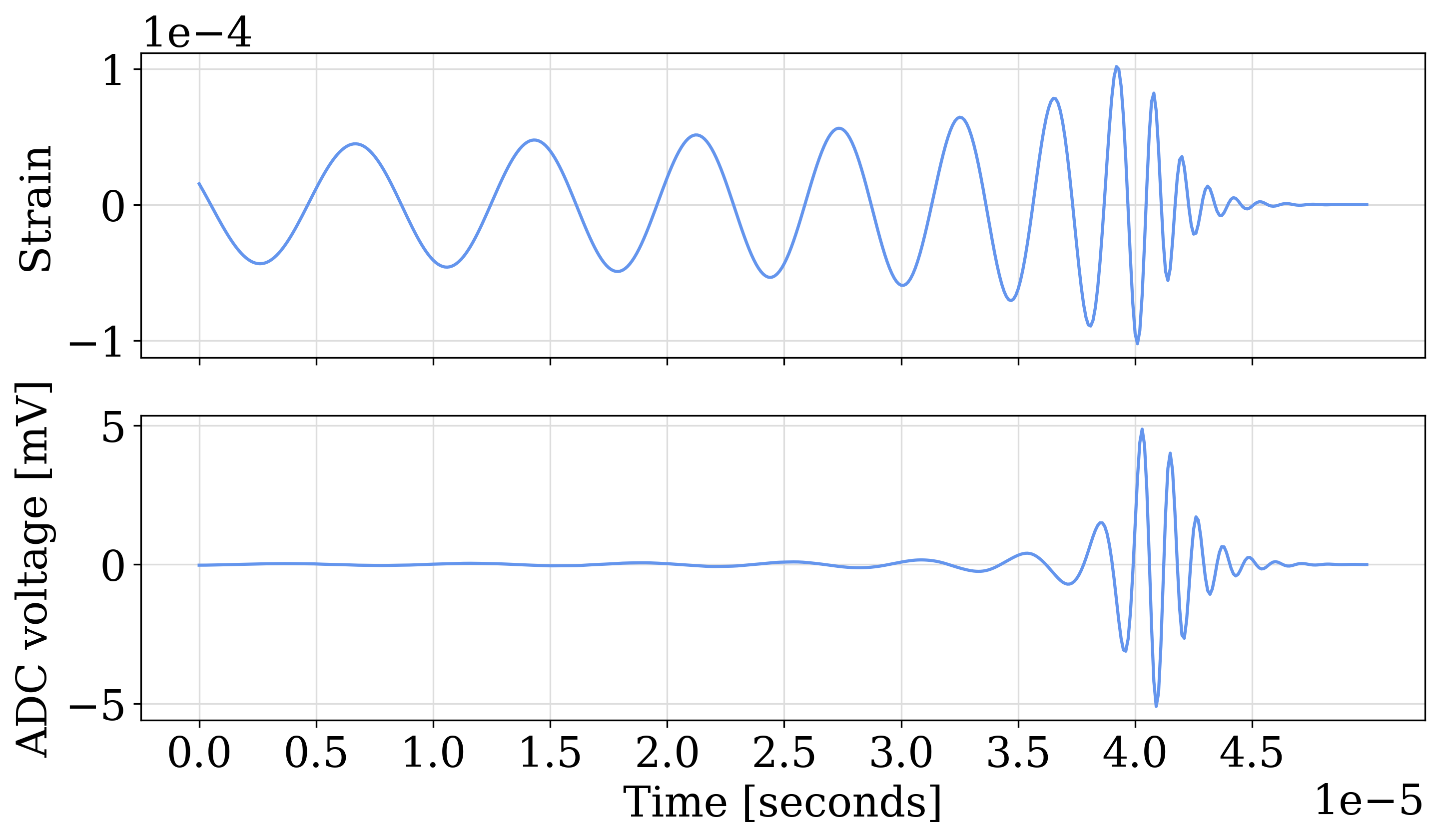}
    \vspace{-0.3cm}
    \caption[Signal template for ABRA-GW]{Signal of a $0.01\,M_{\odot}$ equal mass merger at 100~km distance with the parameters from Table~\ref{tab:template_parameters}. Top: strain from $h^+$ and $h^{\times}$ polarizations added together, band-pass filtered and an irfft taken. Bottom: transformed $h^+$ and $h^{\times}$ polarizations to the detector frame, band-pass filtered and an irfft taken.}
    \label{fig:signal}
    \vspace{-0.5cm}
\end{figure}

After finding the flux on the pickup from COMSOL, the gain resulting from the calibration of the HFGW pickup loop with the HFGW calibration loop were used to convert to mV on the digitizer,
\begin{equation}
    \Phi_{\mathrm{pickup}} \times \mathcal{T}(f)_{\Phi_{\mathrm{pickup}} \rightarrow V_{\mathrm{ADC}}} = V_{\mathrm{ADC}}.
\end{equation}
The transfer function for flux on pickup to voltage on ADC, $\mathcal{T}(f)_{\Phi_{\mathrm{pickup}} \rightarrow V_{\mathrm{ADC}}}$, is found from Eq.~(S1) of Ref.~\cite{2021PRL} dividing the total gain, $ V_{\mathrm{ADC}} / V_{\mathrm{Sig} }$, by $ \Phi_p / I_C \times I_C /  V_{\mathrm{Sig}}$, the mutual inductance between the pickup loop and calibration loop, and the attenuation and current conversion between the signal generator and calibration loop. The results for the two polarizations were then added, an inverse-real FFT was performed, and the waveforms were band-pass filtered from 10\,kHz to 3\,MHz. Figure~\ref{fig:signal} shows the difference between a waveform which was only band-pass filtered in units of strain and one that had the transfer functions applied with units of mV.

\subsection{Gaussian Process Modeling}
Gaussian Process Modeling (GPM) treats the data as a realization of a potentially nontrivial gaussian process, allowing for analyses to replace inference of parametrically specified background models with inference of hyperparameters that specify the kernel of the gaussian process covariance. In choosing a kernel, we aim to specify a covariance model which is sufficiently flexible to accurately model smooth backgrounds while not being so flexible that it can also generate signal-like features, which would result in degraded detection sensitivity.

For the analysis presented here, in which we search for GW-like signals as they appear in the time-domain of ADC voltage measurements, we choose the exponential sine-squared kernel,
\begin{equation}
    K(t,t'| A_\mathrm{GP}, \Gamma, P) = A_\mathrm{GP}\mathrm{exp} \left ( -\Gamma \sin^2\left [\frac{\pi}{P}|t - t'| \right ] \right ).
\end{equation}
This kernel specifies a covariance between time-domain measurement of the background process at times $t$ and $t'$, in terms of the covariance normalization $A_\mathrm{GP}$, the covariance scale $\Gamma$, and the covariance periodicity $P$. This is a safe choice of kernel as our interest is in inspiral signals that chirp, realizing time-varying periodicity, while the period of our kernel is stationary and only allowed to take values in a range which is mismatched with respect to characteristic signal frequencies. We make use of the python package \texttt{George}\cite{2015ITPAM..38..252A} to implement our GPM.

To perform our analysis, we consider detector data at $\mathbf{s}$, where $\mathbf{s}_i$ is the ADC voltage at time $t_i$, and a signal template $\mathbf{h}(d)$, where $\mathbf{h}_i(d)$ is the prediction for the GW signal at time $t_i$ for an inspiral at a distance $d$ from the detector. We then construct the residuals or the signal-subtracted data $\mathbf{r} = \mathbf{s} - \mathbf{h}$, which we assume are described by a mean-zero gaussian process with covariance $\mathbf{K}_{ij} = K(t_i, t_j | A_\mathrm{GP}, \Gamma, P)$. We then calculate the marginalized likelihood of the signal-subtracted data by 
\begin{equation}
    \ln p(\mathbf{r}| d) = - \frac{1}{2} \mathbf{r}^T \mathbf{K} ^{-1} \mathbf{r} - \frac{1}{2} \ln \mathrm{det}\,\mathbf{K} - \frac{n}{2}\ln(2\pi).
\end{equation}
where $n$ is the length of data vector $\mathbf{r}$. Note that this likelihood also depends on the GPM hyperparameters, which we have omitted from equation above for compactness.

We treat this likelihood following standard frequentist techniques, performing maximum likelihood inference to estimate the distance of a putative binary inspiral as well as the GPM hyperparameters. We perform these likelihood maximizations using the \texttt{iminuit} code package~\cite{iminuit}. In addition, we determine 90\% central confidence intervals associated with a $5^\mathrm{th}$ percentile lower limit and $95^\mathrm{th}$ percentile upper limit assuming the asymptotic limit from the Fisher information matrix calculated about the maximum likelihood estimate~\cite{Cowan:2010js}. We also define a test statistic, denoted $\mathrm{TS}$, for an inspiral at distance $d$ given by
\begin{equation}
\mathrm{TS}(d) = -2 \ln \frac{p(\mathbf{r}|d)}{p(\mathbf{r}|\hat{d})}
\end{equation}
where $\hat{d}$ is the maximum likelihood estimate of the signal distance and the probability of the data given the inspiral distance are independently maximized for both $\hat{d}$ and $d$. This test statistic, in the asymptotic limit, can be interpreted as a $\chi^2$-distributed quantity, see \cite{Frate:2017mai,Foster:2021ngm}. We also determine $\mathrm{16}^\mathrm{th}$ percentile lower limits following the power-constrained limit procedure \cite{Cowan:2011an}.

\begin{figure}[!t]
    \centering
    \includegraphics[width=0.45\textwidth]{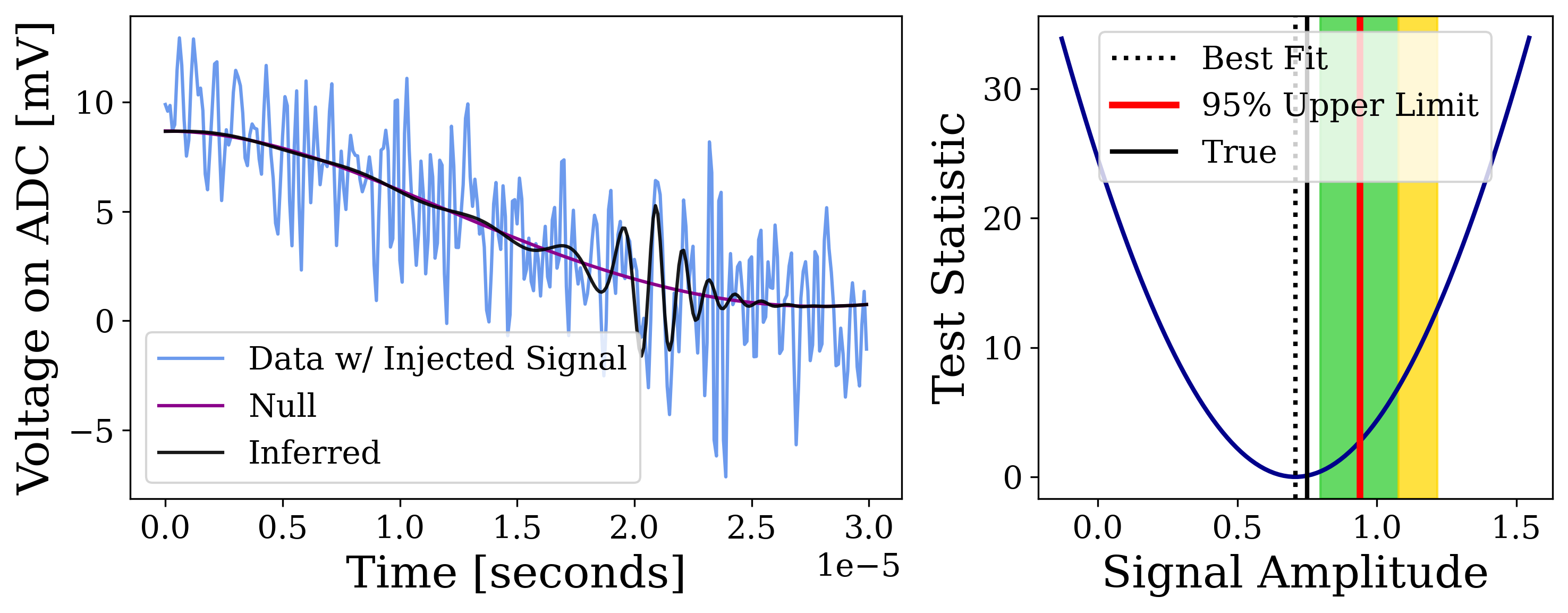}
    \vspace{-0.3cm}
    \caption[Gaussian Process Regression]{Gaussian Process Modeling run on a single segment of data. Left: the ``Null" line was produced by the GPM fit to the background and the ``Inferred" line is that background plus the signal with the best-fit amplitude found from the regression. Right: The profile likelihood is shown in dark-blue with a minimum at the best-fit amplitude value. 90\% confidence with one-sigma bounds (green) and two-sigma bound (yellow).}
    \label{fig:GPExample}
    \vspace{-0.3cm}
\end{figure}

As an example of our analysis, in Fig.~\ref{fig:GPExample}, we show the GPM analysis and TS minimization on one segment of data with an injected signal at a distance of 133 km, which corresponds to $100 \, \mathrm{km} / A$ where $A$ is the amplitude of the signal and has a value of 0.75 for this injection. We repeat this procedure, sliding the time of the inspiral event to inspect the data for a signal which appears over a finite time interval.

\subsection{Binary exclusion}

\begin{figure}[!t]
    \centering
    \includegraphics[width=0.45\textwidth]{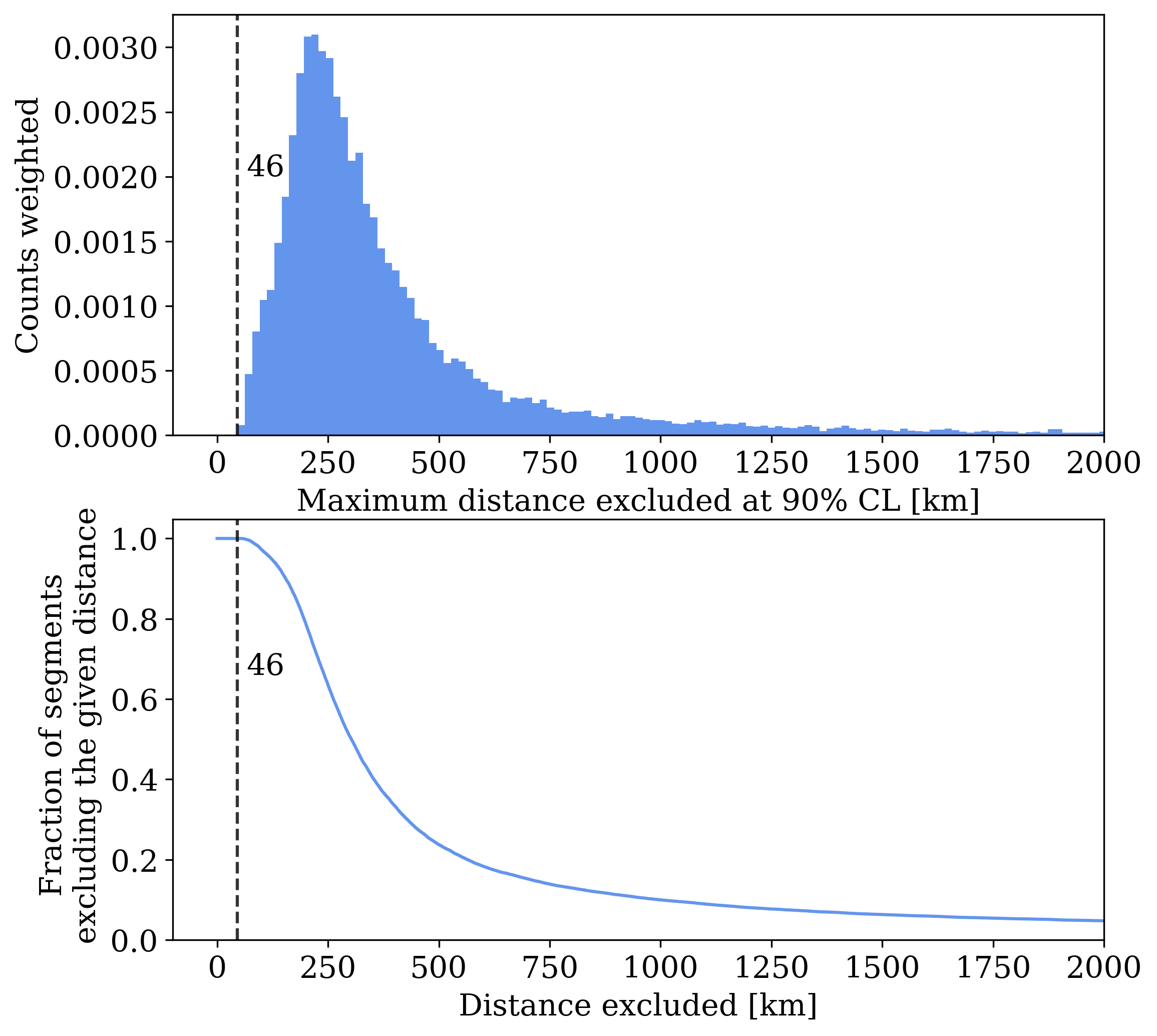}
    \vspace{-0.5cm}
    \caption[Binned exclusion]{Top: weighted distribution of 90\% CL results for the GPM analysis over 10 ms of data. Bottom: fraction of windowed segments which exclude a given amplitude to at least a 90\% CL.}
    \label{fig:Binned}
    \vspace{-0.3cm}
\end{figure}

To determine the exclusion rate on PBHs, we performed the GPM analysis on each windowed segment of data. For each of these segments we collected a best-fit amplitude and an error on that amplitude, which were used to create the 90\% CL for each segment. We found the largest 90\% CL in these results, which gave us the distance of 46.03 km quoted in the main body of this paper. All other segments excluded further distances, meaning the 90\% CL on that distance is the most conservative result over that interval. Figure~\ref{fig:Binned} shows the distribution in 90\% CLs over the windowed segments, along with the fraction of segments which exclude a given distance.

\subsection{Data Stability and Variance}

\begin{figure}[!t]
    \centering
    \includegraphics[width=0.95\linewidth]{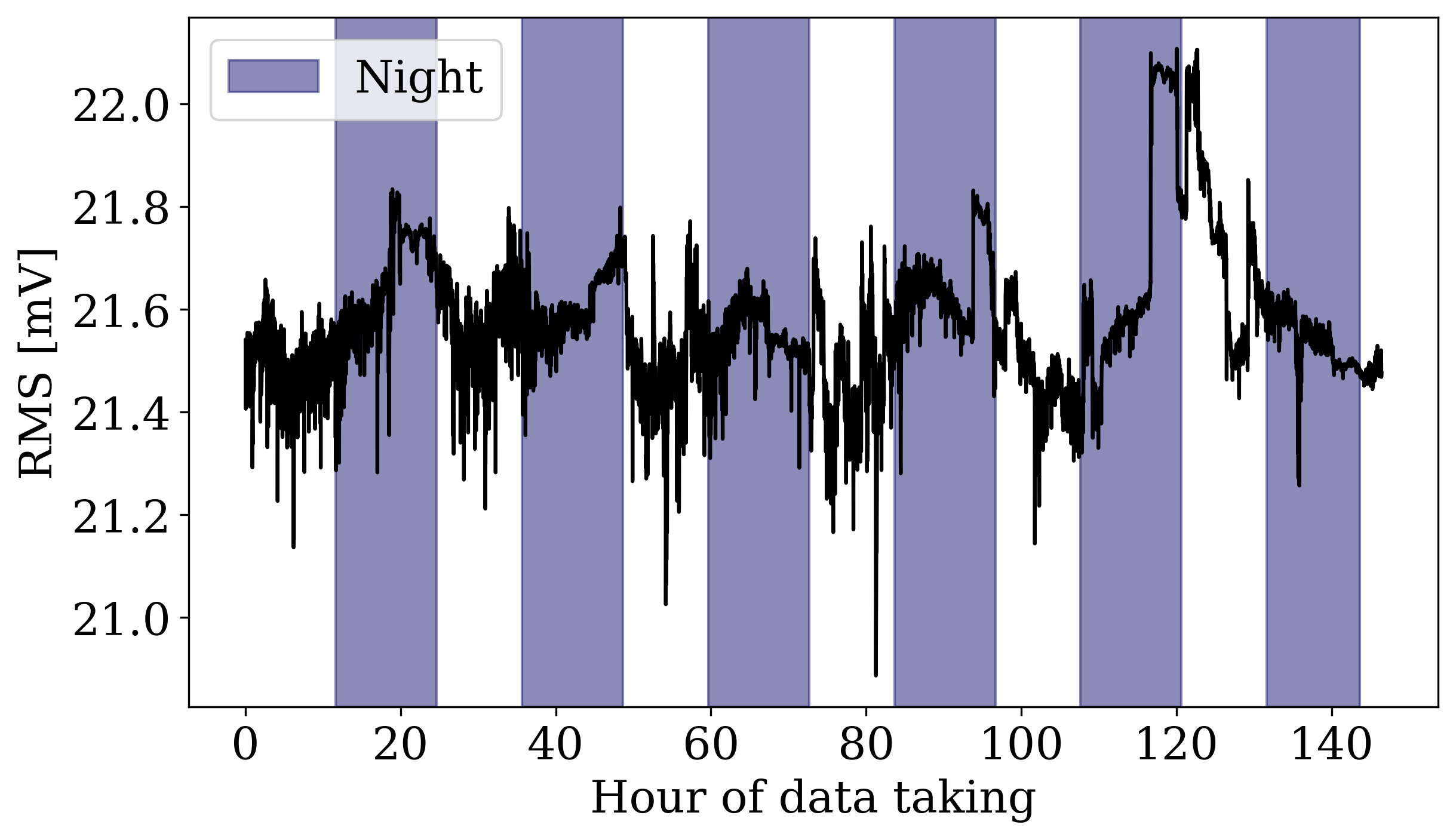}
    \vspace{-0.3cm}
    \caption{Root-mean-square of the variance of the ABRA-GW HFGW data. The portions of data-taking that occurred during the night are marked in purple.}
    \label{fig:enter-label}
    \vspace{-0.3cm}
\end{figure}

Over the course of data taking in December of 2023 we measured the variance on the HFGW data. As can be seen in Fig.~\ref{fig:enter-label}, the RMS of the variance is tightly confined between roughly $20.8$\,mV to $22.1$\,mV, with a vague pattern of noise increasing during the night. We define ``night" for winter in Cambridge MA as the time range from around 17:00 to 7:00. The slight increase in variance during the night is likely resulting from generators turning on in the building or other environmental effects.

\subsection{Effective Current in Toroidal Magnet}

Here we provide the full form of the GW current at the leading order of $ \omega^{2} L^{2} $ appropriate for a toroidal magnet, as originally determined in Ref.~\cite{domcke2024symmetries} including a higher order. The results are divided into three parts: one proportional to $\left[ \Theta(R + a - \rho) - \Theta(R-\rho) \right] \left[ \Theta(H-2z) + \Theta(H+2z) \right]$ describing the volume current, one proportional to $\delta(R + a - \rho) - \delta(R-\rho)$ describing the current on the inside and outside of the volume and one proportional to $\delta(H - 2z) - \delta(H + 2z)$ describing the top and bottom of the magnet. Note the dimensions of the magnet are $H = 120$\,mm and $R = a = 30$\,mm. In the following, we use the shorthand $s_{\theta}$ and $c_{\theta}$ for $\sin \theta$ and $\cos \theta$. We further take $\phi_h = 0$, but $\phi_h$ can be restored by taking $\phi \rightarrow \phi - \phi_h$, which was done when inputting the equations into COMSOL.

\begin{widetext}

    \begin{align}
j_{\rho} e^{i \omega t} =\, &\frac{B_{\max} R  \omega^{2} }{\rho} \left[  \Theta(R+a - \rho) - \Theta(R-\rho) \right] \left[  \Theta(H - 2z) + \Theta(H + 2z) \right]   \\
\times &\Bigg[ \frac{1}{12\sqrt{2}} h^+  
\Big\{ 2z (3 + c_{2\theta_h}) c_{2\phi} + 6 z s_{\theta_h}^2 - \rho c_{\phi} s_{2\theta_h}  \Big\}  + \frac{1}{6\sqrt{2}} h^{\times}  \left( 8 z c_{\theta_h}c_{\phi} - \rho s_{\theta_h}\right) s_{\phi} \Bigg]  \nonumber \\ 
%
%
%
%
-\, & \frac{B_{\max} R \omega^{2} }{\rho} \left[  \Theta(R+a - \rho) - \Theta(R-\rho) \right] \left[  \delta(H - 2z) - \delta(H + 2z) \right]  \nonumber \\
\times & \Bigg[ \frac{1}{12\sqrt{2}}  h^+  \Big\{ (z^2+2\rho^2) (3+c_{2\theta_h})c_{2\phi} + 2zs_{\theta_h}(-4\rho c_{\theta_h}c_{\phi} + 3zs_{\theta_h}) \Big\} + \frac{1}{3\sqrt{2}}  h^{\times}  
 \Big\{ (z^2+2\rho^2) c_{\theta_h} s_{2\phi} - 2z\rho s_{\theta_h} s_{\phi}   \Big\}
\Bigg]  \nonumber
\end{align}

\begin{equation}
\begin{aligned}
j_{\phi} e^{i \omega t} =\, & \frac{B_{\max} R \omega^{2}}{\rho} \left[  \Theta(R+a - \rho) - \Theta(R-\rho) \right]  \left[  \Theta(H - 2z) + \Theta(H + 2z) \right] \\
\times &
 \Bigg[ - \frac{1}{6\sqrt{2}} h^+  \left\{ z (3 + c_{2\theta_h}) c_{\phi} + \rho s_{2\theta_h} \right\}  s_{\phi} 
+  \frac{1}{3\sqrt{2}} h^{\times} \left( z c_{\theta_h} c_{2\phi} + \rho c_{\phi} s_{\theta_h}\right) \Bigg] \\ 
%
%
%
+\, &B_{\max} R \omega^{2} \left[  \delta(R+a - \rho) - \delta(R-\rho) \right] \left[ \Theta(H-2z) + \Theta(H+2z)  \right] \\
\times & \Bigg[ \frac{1}{12\sqrt{2}} h^+   \left\{ z (3 + c_{2\theta_h}) c_{\phi} + \rho s_{2\theta_h} \right\}  s_{\phi} -  \frac{1}{6\sqrt{2}} h^{\times}  \left( z c_{\theta_h} c_{2\phi} + \rho c_{\phi} s_{\theta_h}\right) \Bigg]   \\ 
%
%
%
+\, & \frac{B_{\max} R z \omega^{2} }{\rho} \left[  \delta(R+a -\rho) - \delta(R-\rho) \right] \left[  \delta(H - 2z) - \delta(H + 2z) \right] \\
\times & \Bigg[ \frac{1}{12\sqrt{2}} h^+  \Big\{ 2 \rho s_{2\theta_h} s_{\phi} + z (3+c_{2\theta_h}) s_{2\phi}  \Big\} - \frac{1}{3\sqrt{2}} h^{\times}    (zc_{\theta_h} c_{2\phi}+\rho c_{\phi} s_{\theta_h})
\Bigg] 
\end{aligned}
\end{equation}
%

%
\begin{align}
j_z e^{i \omega t} =\, & \frac{B_{\max} R \omega^{2}}{\rho^2} \left[  \Theta(R+a - \rho) - \Theta(R-\rho) \right] \left[ \Theta(H-2z) + \Theta(H+2z) \right]  \\
\times & \Bigg[ \frac{1}{12\sqrt{2}} h^+   
\Big\{ (z^2 - 2 \rho^2) (3 + c_{2\theta_h}) c_{2\phi}  + 3z \rho c_{\phi} s_{2\theta_h} \Big\} +  \frac{1}{6\sqrt{2}} h^{\times} \Big\{ 4(z^2 - 2 \rho^2) c_{\theta_h} c_{\phi} + 3z \rho s_{\theta_h} \Big\}  s_{\phi} \Bigg] \nonumber \\ 
%
%
%
%
+ \, &\frac{B_{\max} R \omega^{2}}{\rho} \left[  \delta(R+a - \rho) - \delta(R-\rho) \right] \left[ \Theta(H-2z) + \Theta(H+2z) \right] \nonumber \\
\times & \Bigg[ \frac{1}{24\sqrt{2}} h^+ 
\Big\{ (z^2 + 2\rho^2) (3+c_{2\theta_h}) c_{2\phi}  + 2z s_{\theta_h} (- 4\rho c_{\theta_h} c_{\phi} + 3z s_{\theta_h}) \Big\} + \frac{1}{3\sqrt{2}} h^{\times}   \Big\{ (z^2 + 2\rho^2) c_{\theta_h} c_{\phi} - z \rho s_{\theta_h} \Big\} s_{\phi}  \Bigg]  \nonumber
%
%
\end{align}

\end{widetext}

\end{document}